\documentclass[twocolumn,superscriptaddress,floatfix,aps]{revtex4}
\usepackage{graphicx}
\usepackage{amsmath}
\usepackage{amssymb}
\usepackage{latexsym}
\usepackage{array}
\usepackage{amsfonts}
\usepackage{mathrsfs}
\usepackage{verbatim}

\begin{document}

\title{A quantum heuristic algorithm for traveling salesman problem}

\author{Jeongho Bang}
\affiliation{Center for Macroscopic Quantum Control \& Department of Physics and Astronomy, Seoul National University, Seoul, 151-747, Korea}
\affiliation{Department of Physics, Hanyang University, Seoul 133-791, Korea}

\author{Junghee Ryu}
\affiliation{Institute of Theoretical Physics and Astrophysics, University of Gda\'{n}sk, 80-952 Gda\'{n}sk, Poland}
\affiliation{Department of Physics, Hanyang University, Seoul 133-791, Korea}

\author{Changhyoup Lee}
\affiliation{Centre for Quantum Technologies, National University of Singapore, 3 Science Drive 2, Singapore 117543}
\affiliation{Department of Physics, Hanyang University, Seoul 133-791, Korea}

\author{Seokwon Yoo}
\affiliation{Department of Physics, Hanyang University, Seoul 133-791, Korea}

\author{James Lim}
\affiliation{Department of Physics, Hanyang University, Seoul 133-791, Korea}

\author{Jinhyoung Lee}
\affiliation{Center for Macroscopic Quantum Control \& Department of Physics and Astronomy, Seoul National University, Seoul, 151-747, Korea}
\affiliation{Department of Physics, Hanyang University, Seoul 133-791, Korea}
\affiliation{School of Computational Sciences, Korea Institute for Advanced Study, Seoul 130-722, Korea}
\received{\today}

\begin{abstract}
We propose a quantum heuristic algorithm to solve a traveling salesman problem by generalizing Grover search. Sufficient conditions are derived to greatly enhance the probability of finding the tours with the cheapest costs reaching almost to unity. These conditions are characterized by statistical properties of tour costs and shown to be automatically satisfied in the large number limit of cities. In particular for a continuous distribution of the tours along the cost we show that the quantum heuristic algorithm exhibits the quadratic speedup over its classical heuristic algorithm.
\end{abstract}

\maketitle

\newcommand{\bra}[1]{\left<#1\right|}
\newcommand{\ket}[1]{\left|#1\right>}
\newcommand{\abs}[1]{\left|#1\right|}
\newcommand{\expt}[1]{\left<#1\right>}
\newcommand{\braket}[2]{\left<{#1}|{#2}\right>}
\newcommand{\commt}[2]{\left[{#1},{#2}\right]}

\section{Introduction}

The quantum computation are potentially more powerful than its classical counterparts \cite{Nielsen99}. In particular some quantum algorithms enabling the dramatic speedup has opened new possiblities of quantum computation to solve hard problems \cite{Shor97,Grover97,Harrow09,Childs09}. It is usually considered that the hard problems have so many candidate solutions that it is hard (or practically impossible) to search the exact solution among them quickly. Such hard problems are usually classified as ``non-deterministic polynomial (NP) class'' in complexity theory \cite{Garey79}. Thus researchers in this field have expected a succession of quantum agorithm to solve the NP problems, depending on {\em quantum superposition} which enables to investigate simultaneously all the states. However the development of quantum algorithms has been rather stagnant, with any remarkable quantum algorithms having not been discovered in the last few years \cite{Shor03}, and it is still unclear that all NP problems can efficiently be solved by quantum algorithm.

On the other hand, in computational science, heuristic approach is one of the most efficient ways to solve the NP problems. In heuristic approach, the problem is usually simplified at the cost of accuracy, and thus it may fail to find the exact solution \cite{Hogg00}. Heuristic algorithm, instead, enables us to gain computational speedeup by finding the approximate solution, near to the exact one, so quickly. Therefore it is natural that current research in quantum algorithm has been directed towards using the heuristic approach \cite{Hogg98, Hogg00,Trugenberger02}. However, it also has not been elucidated thoroughly that the heuristic approach is {\em always} helpful for speedup when we deal with the problem in a quantum-mechanical way. The present work is actually motivated by the goal of investigating this issue.

As aforementioned, it is generally known that the heuristic approach is useful to solve the hard problems. However, in the case of using the quantum system, this remains valid only for the search problem \cite{Hogg98,Grover99}. In addition, the heuristic algorithms are problem-specific: There is no guarantee that a heuristic procedure used for one problem will be applicable to others \cite{Kirkpatrick83,Znidaric06,Farhi08}. Therefore we consider a particular set of problems, called ``NP-complete'', because of the property that {\em all NP problems can efficiently be mapped into a problem in this class} \cite{Garey79}. Typical examples belonging to the NP-complete class are a traveling salesman problem (TSP) and an exact cover \cite{Applegate06}. We consider exclusively the TSP in this paper. Thus, if a effective heuristic algorithm for TSP were found, it could be mapped into a procedure for solving all NP problems even though problem-specific.

In TSP, a salesman travels $n$ cities by visiting each city once and only once, and returning to the starting city. The problem is to find the lowest-cost tour among $N=(n-1)!$ possible tours, where the symbol ``$!$'' stands for the factorial. Throughout the paper, we assume that the oracle answers only the overall cost for the given tour when queried, hiding the details on the costs of city pairs. This assumption is made in order to compare the performances of algorithms under the equal conditions, neglecting any improvement by structural information (see Ref.~\cite{Karp82} for such an improvement). With such an oracle of the tour costs, a classical exact algorithm is to query the costs of all the tours until finding the cheapest. As the cheapest cost is unknown, contrary to the data search where the target is known \cite{Grover97}, the search should be done over all the tours. This demands a huge number $(n-1)! \simeq e^{n \ln n + n}$ of queries, independent of whether the algorithm is deterministic or probabilistic.

Many quantum approaches for the TSP (or equivalent to this problem) have been studied \cite{Cerny93,Hogg00,Farhi01,Trugenberger02,Martonak04}. In particular the quantum annealing method was investigated for the TSP \cite{Martonak04,Trugenberger02}. It has been discussed, however, that finding a ground state becomes extremely difficult as increasing the size of system (or equivalently the number of cities) \cite{Znidaric06,Farhi08}. It is thus unclear that the quantum annealing can be applied to the extremely large number of cities. On the other hand, one may convert a TSP to the problem of finding a minimum for a given cost function, called ``a minimum search.'' A quantum algorithm to obtain an exact solution of minimum search was proposed, based on Grover search, which works in ${\cal O}(\sqrt{N})$ queries, where $N$ is the number of elements (or tours in TSP) \cite{Durr96}. This is the quadratic speedup of the classical exact algorithm earlier, as $\sqrt{N} \simeq e^{\frac{1}{2} n \ln n + \frac{1}{2} n}$. 

We instead consider a classical heuristic algorithm which is adapted to the circumstance: The tours along the cost $c$ are distributed, close to a normal distribution with the average $\bar{c}$ and the standard deviation $\Delta c$ in the limit of the large number of cities, $n \rightarrow \infty$. This circumstance has commonly been considered in the literatures \cite{Applegate06,Hogg00}. (The statistical information can easily be estimated by sampling a small number of tours \cite{Beardwood59,Burgess89}.) Taking an approximate solution to be a tour whose cost is less than $\Delta c$, the probability $f$ of finding a solution is asymptotically given, 
\begin{equation}
\label{eq:chap}
f \simeq e^{-\frac{3}{2} n - \frac{1}{2} \ln n}.
\end{equation}
Then the classical heuristic algorithm is following: Query $1/f$ tours that are randomly chosen and find the cheapest one among those tours. Then one solution is {\em likely} to be found in ${\cal O}(1/f)$ queries. Note that this classical heuristic algorithm is much faster than the exact quantum algorithm by removing the logarithm in the exponent.

In this paper, we propose a quantum heuristic algorithm for TSP by generalizing Grover operation. Sufficient conditions are derived under which the algorithm can greatly enhance the probability of finding approximate tours with the cheapest costs, reaching almost to unity. These conditions are characterized by statistical properties of tour costs and shown to be automatically satisfied in the large number limit of cities. The present quantum heuristic algorithm is shown to work in ${\cal O}(1/\sqrt{f})$ queries, a quadratic speed up of the classical heuristic algorithm with its order ${\cal O}(1/f)$ in Eq.~(\ref{eq:chap}).

\section{Quantum traveling salesman problem}

To represent the tours quantum-mechanically, an orthonormal basis set of $N$ states are appropriate and they are fully discriminated by a von-Neumann orthogonal measurement. We assume that a quantum register is of $N$ dimension for a sake of simplicity and its basis states $\ket{T}$ respectively correspond to the tours $T$. 

\begin{figure}[t]
\begin{center}
\includegraphics[width=0.45\textwidth]{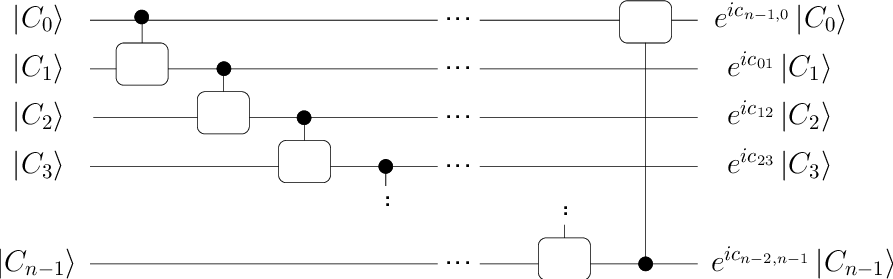}
\caption{Simple quantum circuit for cost oracle. The box coupled with another channel stands for the conditional phase operation $\hat{P}_{jk}$. The cost oracle can efficiently be constructed with $n$ number of conditional phase operations.}
\label{fig:cost-oracle}
\end{center}
\end{figure}

An oracle operation $\hat{C}$ is introduced to answer the tour cost, when applied, such that
\begin{eqnarray}
\label{eq:coo}
\hat{C} \ket{T} = e^{i \phi(T)} \ket{T},
\end{eqnarray}
where $\phi(T)$ stands for the overall cost, given tour $T$, and it is called a cost phase. Here, every cost phase is defined in between $0$ and $2\pi$ by scaling the tour costs. The scaling can easily be made once obtaining the strictly lower and upper bounds of the tour costs \cite{Trugenberger02}. {\em It is remarkable that the oracle constructed will not expose any details such as the city-pair costs but just the overall tour costs}. In order to construct the cost oracle, it is convenient to define a tour state as 
\begin{eqnarray}
\ket{T} = \ket{C_0}\otimes\ket{C_1}\otimes\cdots\otimes\ket{C_{n-1}},
\label{eq:t-st}
\end{eqnarray}
where $\ket{C_k}$ stands for the k-th visiting city, and $\ket{C_0}$ is both the starting city and finally visiting city. We note here that Hilbert space spanned by all possible $\ket{T}$ states defined as Eq.~(\ref{eq:t-st}) is actually given as $n^n$, however, we consider only $N=(n-1)!$ subspace for valid tour states with the condition $C_j \neq C_k$ for different $j$ and $k$ to each other \cite{Goswami04}. Such a coding scheme is often met in practical applications (See Refs. \cite{Zanardi97, Lanyon08}). The conditional operation $\hat{P}_{jk}$ is also introduced for cost oracle, such that $\hat{P}_{jk}\ket{C_j}\ket{C_k} = e^{i c_{jk}}\ket{C_j}\ket{C_k}$, where $c_{jk}$ denotes the cost from $j$-th to $k$-th visiting city. Since this conditional operation $\hat{P}_{jk}$ is efficiently be decomposable to two-qubit and/or Hadamard gates~\cite{Schuch03}, we can finally construct a cost oracle $\hat{C}$ with relatively small computational efforts by using the $\hat{P}_{jk}$ (as depicted in Fig.~\ref{fig:cost-oracle}).

\section{Generalized Grover operation}

We generalize Grover's data search \cite{Grover97} and apply its generalization to amplifying the probability for approximate solutions. Before the generalization, we briefly discuss the Grover operator $\hat{G}_0$ in an alternative perspective. Applying $\hat{G}_0$ iteratively leads an initial state $\ket{\psi_0}$ in an equal superposition of the states representing the data toward a target $\ket{\psi_1}$ which is a specific data desired to be found. [G.1] It is a composition of two reflections: 
\begin{equation}
 \hat{G}_0 = \hat{I}_{\psi_0} \hat{I}_{\psi_1},
\end{equation}
where $\hat{I}_{\psi_i} \ket{\psi_i} = - \ket{\psi_i}$ and $\hat{I}_{\psi_i} \ket{\psi_i^{\perp}} = \ket{\psi_i^{\perp}}$, $\forall \ket{\psi_i^\perp}$ orthogonal to $\ket{\psi_i}$. [G.2] Small overlap between $\psi_i$: $ \sqrt{f} \ll 1$, where $\sqrt{f} = \braket{\psi_0}{\psi_1}$ is assumed to be a real number for a sake of simplicity. [G.3] $\hat{G}_0$ is a rotation of angle $2\sqrt{f}$ except a global phase, up to the first order of $\sqrt{f}$, on the plane spanned by $\ket{\psi_0}$ and $\ket{\psi_1}$: Letting $\ket{\psi_1^\perp} = (\ket{\psi_0} - \sqrt{f} \ket{\psi_1})/\sqrt{1-f}$,
\begin{eqnarray}
\hat{G}_0  \approx - \exp(i 2\sqrt{f} \hat{\sigma}_y) \oplus \hat{\openone}_c
\end{eqnarray}
where $\hat{\openone}$ is an identity operator, $\hat{\sigma}_y$ a Pauli spin operator in the subspace spanned by $\{\ket{\psi_1}, \ket{\psi_1^\perp}\}$, and $\hat{\openone}_c$ is an identity operator defined in the complementary subspace. Note that $\ket{\psi_0} \simeq \exp(i \sqrt{f} \hat{\sigma}_y) \ket{\psi_1^\perp}$. [G.4] The number of iterations from $\ket{\psi_0}$ to the target $\ket{\psi_1}$ is $R=\left[(\pi/2 - \sqrt{f})/2\sqrt{f}\right]$, where $[x]$ is the nearest integer to $x$, that is, $\hat{G}_0^R \ket{\psi_0} \simeq \ket{\psi_1}$.

We propose a generalized Grover operator, defined by
\begin{eqnarray}
\label{eq:ggo}
 \hat{G} = - \hat{I}_{\psi_0} \hat{C},
\end{eqnarray}
where $\hat{C}$ is the cost oracle in Eq.~(\ref{eq:coo}) and $\hat{I}_{\psi_0} \ket{\psi_0} = - \ket{\psi_0}$ while leaving other orthogonal states invariant. Here, the initially prepared state $\ket{\psi_0}$ of the quantum register is given by
\begin{eqnarray}
\ket{\psi_0} = \frac{1}{\sqrt{N}} \sum_{T} \ket{T},
\label{eq:init_st-2q}
\end{eqnarray}
where the sum is over the all $N$ of tours. The generalized Grover operator $\hat{G}$ in Eq.~(\ref{eq:ggo}) consists of not two reflections but a single reflection and the cost oracle. This is the distinct feature of the generalized Grover operator, different from those investigated in Ref.~\cite{Grover98, Biham00}. 

The definition of $\hat{G}$ is made by observing two simple cases. In each case, the tours are divided into several groups according to their costs. The $j$-th group contains the number $N_j(c)$ of tours and their costs $c$ are all to be encoded to a certain cost phase $\phi_j$. The first case is two groups: One of $N_1$ tours having $\phi_1=0$, and the other of $N_2=N-N_1$ tours having $\phi_2=\pi$. Assuming $N_1 \ll N$, $\hat{G}$ is directly approximated to the Grover operator $\hat{G}_0$, upto the first order of $\sqrt{f}=\sqrt{N_1/N}$. We can then find the tours of $N_1$ group with near unity probability. 

The other simple case is that the tours are divided into four groups, each of which has the cost phase $\phi_j = j \pi/2$ for $j=0,1,2,3$, respectively. Assume further that $N \simeq N_2 \gg N_j$ for $j \ne 2$. It is convenient to define four states $\ket{\phi_j}$ by
\begin{eqnarray}
\ket{\phi_j} = \frac{1}{\sqrt{N_j}}\sum_{T \in S_j} \ket{T},
\label{eq:cost-ph_st}
\end{eqnarray}
where $S_j$ denotes the $j$-th group, and hence the sum is over the tours $T$ in $S_j$. The four states form an orthonormal basis $\{ \ket{\phi_j} \}$, which spans a four dimensional subspace. In particular, the initial state $\ket{\psi_0}$ is expanded in terms of the basis as
\begin{eqnarray}
\ket{\psi_0} = \sum_{j=0}^3 \sqrt{f_j} \ket{\phi_j},
\end{eqnarray}
where $f_j = N_j/N$. The subspace spanned by $\{\ket{\phi_j}\}$ remains invariant under the transformations by the cost oracle $\hat{C}$ and the reflection $\hat{I}_{\psi_0}$: $\hat{C} \ket{\phi_j} = e^{i \phi_j} \ket{\phi_j}$ and $\hat{I}_{\psi_0} \ket{\phi_j} = \ket{\phi_j} - 2\sqrt{N_j/N} \ket{\psi_0} = \sum_{j} c_j\ket{\phi'_j}$ with some real numbers $c_j$. As long as staying in the subspace, both operators can be represented in terms of $\{\ket{\phi_j}\}$. In particular, the cost oracle $\hat{C}$ is reduced and equivalent to
\begin{eqnarray}
\hat{C}_s = - \hat{\openone}_s + \sum_{j} (1+e^{i \phi_j}) \ket{\phi_j} \bra{\phi_j},
\end{eqnarray}
where $\hat{\openone}_s = \sum_j \ket{\phi_j} \bra{\phi_j}$ is an identity operator in the subspace. The necessary condition similar to [G.2] is cast in the form of
\begin{eqnarray}
\left| \bra{\psi_0} ( \hat{C}_s + \hat{\openone}_s ) \ket{\psi_0}\right| \approx 0,
\label{eq:solc}
\end{eqnarray}
which is equivalent to $\sqrt{f_j} \ll 1$ for $j \ne 2$. We may regard $\hat{C}_s$ as the cost oracle, omitting the subscript $s$, without loss of any generality. Within the subspace, the generalized Grover operator is approximated, upto the first order of $\sqrt{f_j}$ for $j \ne 2$, as
\begin{eqnarray}
\hat{G} &\simeq& -\ket{0} \bra{0} - \ket{\pi} \bra{\pi} - i \ket{\pi/2} \bra{\pi/2} + i \ket{3\pi/2} \bra{3\pi/2} \nonumber \\
&& - 2 \sum_{j \ne 2}^3 \sqrt{f_j} \left( \ket{j\pi/2} \bra{\pi} - e^{j \pi/2} \ket{\pi} \bra{j \pi/2} \right).
\end{eqnarray}
It can be seen that $\hat{G}$ is a transformation between $\ket{\pi}$ and the other basis states. On the other hand, the fourth power
\begin{eqnarray}
\hat{G}^4 \approx \exp(i 8 \sqrt{f_0} \hat{\sigma}_y) \oplus \hat{\openone}_c
\end{eqnarray}
is a rotation of angle $8\sqrt{f_0}$ on the plane spanned by $\ket{\pi}$ and $\ket{0}$, leaving the other two states $\ket{\pi/2}$ and $\ket{3\pi/2}$ unchanged. Here $\hat{\sigma}_y$ is a Pauli spin operator in $\ket{0}$-$\ket{\pi}$ subspace and $\hat{\openone}_c$ is an identity operator in $\ket{\pi/2}$-$\ket{3\pi/2}$ subspace. It is remarkable that, in our approach, we do not have to generalize the ``phase matching condition'' \cite{Long99}. Noting $\ket{\psi_0} \approx \sqrt{-\hat{G}} \ket{\pi}$, then $(\pi/2-\sqrt{f_0})/8\sqrt{f_0}$ iterations of $\hat{G}^4$ approximately transform
\begin{eqnarray}
 &&\ket{\psi_0} = \sqrt{f_0} \ket{0} + \sqrt{f_1} \ket{\pi/2} + \sqrt{f_2} \ket{\pi} + \sqrt{f_3} \ket{3\pi/2} \nonumber \\
&&\longrightarrow \sqrt{f_0+f_2} \ket{0} + \sqrt{f_1} \ket{\pi/2} + \sqrt{f_3} \ket{3\pi/2}.
\end{eqnarray}
It is clearly seen that the probability of finding $\ket{0}$ is greatly enhanced as $f_0+f_2 \simeq 1 \gg f_{1,3}$, reminding of $f_2 \gg f_0,f_1,f_3$. We note that the number of iterations needs to be a multiple of 4 to exclude the possibility of leaking the probability to the other tours. One might be concerned about the case that $f_{1,3} \ge f_0$ but this does not significantly alter our main result as $f_{1,3} \ll f_2$.

This procedure is generalized straightforwardly to $2M$ groups of tours, $j$-th group having phase $\phi_j = j \pi/M$, where $j=0,1,...,2M-1$. The arguments similar to the case of $4$ groups can be applied now with $2M$ dimensional subspace. Then, the $2M$-th power of $\hat{G}$ is given by, up to the first order of $\sqrt{f_j}$,
\begin{eqnarray}
\hat{G}^{2M} \approx \exp(i 4M \sqrt{f_0} \hat{\sigma}_y) \oplus \hat{\openone}_c,
\end{eqnarray}
where $\sum_k \exp(i 2\pi kj/2M) = 2M \delta_{j0}$ is used and $\hat{\sigma}_y$ is defined in $\ket{0}$-$\ket{\pi}$ subspace, the same as the cases of $M=1,2$. In order to obtain the almost-unit probability of finding $\ket{0}$, the number of iterations of $\hat{G}^{2M}$ is given by $(\pi/2-\sqrt{f_0})/4M \sqrt{f_0}$: The number of iterations $R$ of $\hat{G}$ or the number of queries becomes
\begin{equation}
\label{eq:nqqha}
 R = \left[ \frac{\pi/2-\sqrt{f_0}}{2 \sqrt{f_0}} \right].
\end{equation}
We note that an accumulated error, for very large $R$, is order of ${\cal O}(\sqrt{f_0})$, which is {\em negligible}.

We analyzed that the present algorithm works in which the discretization of the tours into $2M$ groups is admitted and $N(\pi) \gg N(\phi_j)$ for all the other phases $\phi$. In other words, {\em if the cost phases $\phi_j=0$ and $\phi_j=\pi$ are well defined and their tour population ratio $f_0/f_\pi$ is much less than unity, the presented algorithm can be applied to solving TSP}.

\section{Validity for continuous model}

We investigate applicability of our algorithm in a continuous tour-cost model with its statistical properties. In other to calculate the total iterations, a Gaussian model is used as a typical example \cite{Applegate06,Hogg00}. To begin with, we define the density of tours $\nu(c)$ taking a given cost $c$, instead of the number of tours $N_j(c)$ in a discretized $j$-th group. The density of tours $\nu(c)$ is defined, in the continuum limit of $n \rightarrow \infty$, by
\begin{eqnarray}
\label{eq:dot1}
 \nu(c) \equiv \sum_T \delta(c(T)-c),
\end{eqnarray}
where $c(T)$ is the overall cost of the given tour $T$ and $\delta(c)$ is a Dirac delta function. We note here that in the continuum limit the initial state $\ket{\psi_0}$ is rewritten as
\begin{eqnarray}
 \ket{\psi_0} = \int_0^{2\pi} d \phi \sqrt{\frac{\nu(\phi)}{N}} \ket{\phi},
\end{eqnarray}
where the states $\ket{\phi}$ is defined by
\begin{eqnarray}
\ket{\phi} \equiv \frac{1}{\sqrt{\nu(\phi)}}\sum_{T} \delta(\phi - \phi(T)) \ket{T}. 
\end{eqnarray}
Such a definition of $\ket{\phi}$ is originated from the Eq.~(\ref{eq:cost-ph_st}). Here, these states $\ket{\phi}$ form the unnormalized basis, satisfying $\braket{\phi'}{\phi} = \delta(\phi' - \phi)$.

Then, we investigate the statistical properties of tour costs, which are the cost average $\overline{c}$ and the standard deviation $\Delta c$. The $\overline{c}$ and $\Delta c$ are easily estimated, assuming that city-pair costs $c_{jk}$ are randomly chosen in the interval of $[c_1,c_2]$ with an identical probability density distribution $P(c_{jk}) = 1/(c_2 - c_1)$. A tour cost $c(T)$ of $n$ cities is in the interval $[nc_1,nc_2]$. The $\overline{c}$ is given by
\begin{eqnarray}
\label{eq:afm}
 \overline{c} = \frac{1}{(n-1)!}\sum_T c(T) = \frac{1}{n-1} \sum_{j=0}^{n-1} \sum_{k \ne j}^{n-1} c_{jk}.
\end{eqnarray}
Here the last equality holds as every $c_{jk}$ appears $(n-2)!$ times in all tours and  $\sum_T c(T) = (n-2)! \sum_{jk}' c_{jk}$, where the sum runs over $k \ne j$. Noting that the ensemble average of every single pair cost $c_{jk}$ is equal to $(c_2-c_1)/2$, the cost average $\overline{c}$ is estimated to
\begin{eqnarray}
 \overline{c}_\mathrm{est} = n \frac{c_2-c_1}{2}.
 \label{eq:sp1}
\end{eqnarray}
The average of the second moment is given by
\begin{eqnarray}
\label{eq:asm}
 \overline{c^2} &=& \frac{1}{(n-1)!} \sum_T c(T)^2 \nonumber \\
&=& \frac{(n-2)!}{(n-1)!} \sum_{j_1 \ne j_2}^{n-1} c_{j_1j_2}^2 + \frac{2 (n-3)!}{(n-1)!} \sum_{j_1 \ne j_2 \ne j_3}^{n-1} c_{j_1j_2} c_{j_2j_3} \nonumber \\
&& + \frac{(n-3)!}{(n-1)!} \sum_{j_1 \ne j_2 \ne j_3 \ne j_4}^{n-1} c_{j_1j_2} c_{j_3j_4},
\end{eqnarray}
where the last equality holds similarly to Eq.~(\ref{eq:afm}). The ensemble average of every $c_{j_1 j_2}^2$ is equal to $(c_2^3 - c_1^3)/3(c_2-c_1)$. On the other hand, every $c_{j_1j_2}c_{j_3j_4}$ is averaged to the simple product of the two ensemble averages of $c_{j_1j_2}$ and $c_{j_3j_4}$, i.e. $[(c_2-c_1)/2]^2$, $(j_1,j_2) \ne (j_3,j_4)$. The second moment is estimated to
\begin{eqnarray}
 \overline{c^2}_\mathrm{est} = n \frac{c_2^2+c_2c_1+c_1^2}{3} + (n^2 -n) \frac{(c_2-c_1)^2}{4}.
\end{eqnarray}
The estimated variance is then given as
\begin{eqnarray}
 \Delta c_\mathrm{est}^2 = \overline{c^2}_\mathrm{est} - \overline{c}_\mathrm{est}^2 = n \frac{c_2^2+10c_2c_1+c_1^2}{12}.
\label{eq:sp2}
\end{eqnarray}
Then the statistical properties of tour costs, the Eq.~(\ref{eq:sp1}) and Eq.~(\ref{eq:sp2}), are briefly represented as
\begin{eqnarray}
\frac{\Delta c}{\overline{c}} = {\cal O}(\frac{1}{\sqrt{n}}).
\label{eq:sp_f}
\end{eqnarray}
The distribution $\nu(c)$ is transformed to $\nu(\phi)$ with respect to the cost phase $\phi$, by the relation $\phi(T) = 2 \pi c(T)/(nc_2 - nc_1)$. It is notable that the average cost phase $\bar{\phi} = \pi$ and the standard deviation $\Delta \phi \sim 1/\sqrt{n}$. {\em In the limit of $n \rightarrow \infty$, the tours population becomes concentrated to the average $\bar{\phi}=\pi$ as $\Delta \phi \rightarrow 0$.} 

We shall finally investigate the applicability of our algorithm to the continuous tour-cost distribution of $n \rightarrow \infty$. Remind that in our approach the crucial conditions are that (a) the two cost phases of $\phi=0$ and $\phi=\pi$ are well defined and (b) their population ratio $f_0/f_\pi$ is much less than unity. The condition (a) is rephrased as the existence of $\eta \ll 1$ such that 
\begin{eqnarray}
\label{eq:conda1}
 \hat{C} \int_{-\eta/2}^{\eta/2} d \phi \sqrt{\frac{\nu(\phi)}{N}} \ket{\phi} &\approx& e^{i 0} \sqrt{f_0} \ket{0} \\
\label{eq:conda2}
 \hat{C} \int_{-\eta/2}^{\eta/2} d \phi \sqrt{\frac{\nu(\pi+\phi)}{N}} \ket{\pi+\phi} &\approx& e^{i \pi} \sqrt{f_\pi} \ket{\pi},
\end{eqnarray}
where $\hat{C}$ is the cost oracle. If $\Delta \phi \ll \eta \ll 1$, it is clearly seen that Eq.~(\ref{eq:conda2}) holds as $\nu(\phi) \approx N \delta(\phi-\pi)$. For the phase $0$, noting $\eta \ll 1$, we expand $\hat{C} \ket{\phi}$ in Eq.~(\ref{eq:conda1}) upto the first order, {\em i.e.} $\hat{C} \ket{\phi} \simeq \ket{\phi} + i \phi \ket{\phi}$. The first order is negligible as
\begin{eqnarray}
\label{eq:vcfcc}
 \frac{\int_{-\eta/2}^{\eta/2} d\phi \phi^2 \nu(\phi)}{\int_{-\eta/2}^{\eta/2} d\phi \nu(\phi)} \approx \frac{\eta^2}{12} \ll 1.
\end{eqnarray}
Thus, the condition (a) holds for both phases $0$ and $\pi$ if $\Delta \phi \ll \eta \ll 1$. By using Eq.~(\ref{eq:sp_f}), the sufficient condition is rewritten as
\begin{equation}
1 \ll \frac{\eta}{\Delta \phi} \ll \frac{1}{\Delta \phi} = {\cal O}(\sqrt{n}). 
\end{equation}
This is also sufficient for the condition (b), as the most of tours are concentrated around the average $\phi=\pi$ within a few of $\Delta \phi$. Thus the conditions (a) and (b) are characterized by the statistical properties in a continuous tour-cost model. 

We shall now investigate the total iteration (or equivalently queries) of the generalized Grover operator in the continuous tour-cost model. To do this we consider a Gaussian density of tours $\nu(c)$ near the lowest costs
\begin{eqnarray}
\label{eq:dot2}
 \nu(c) = N \frac{\nu_0}{\Delta c} \exp\left[- \frac{(c-\bar{c})^2}{2 \Delta c^2}\right],
\end{eqnarray}
where $\overline{c}$ is the average of tour costs and $\Delta c$ the standard deviation: $\overline{c} = \int dc~ c \nu(c)/N$ and $\Delta c^2 = \int dc~ (c-\overline{c})^2 \nu(c)/N$. The normalization factor $\nu_0$ is determined such that $\int dc~ \nu(c) = N$, the total number of tours. Then $f_0$ is approximately given as
\begin{equation}
f_0 = \frac{1}{N} \int_{-\eta/2}^{\eta/2} d\phi~ \nu(\phi) \simeq e^{- \frac{3}{2} n - \frac{1}{2} \ln n},
\label{eq:f0}
\end{equation}
which is exactly same as $f$ in Eq.~(\ref{eq:chap}) in classical heuristic algorithm. Nevertheless, total query is ${\cal O}(1/\sqrt{f_0})$, as seen in Eq.~(\ref{eq:nqqha}), {\em leading to the quadratic speedup over the classical heuristic algorithm}.

Before closing, we indicate that our algorithm enhance the probability of finding the lowest-cost tours together with the highest-cost tours. To see this, recall that $f_0$ is given as the number of tours whose cost phases are from $-\eta/2$ to $\eta/2$, as in Eq.~(\ref{eq:f0}). Here the cost phases between $-\eta/2$ and $0$ are actually the highest costs in $[2\pi-\eta/2, 2\pi]$. This situation is caused by periodicity of the cost phases, {\em i.e.} $e^{-i\epsilon/2}\ket{T} = e^{i(2\pi-\epsilon/2)}\ket{T}$. In this case, for very small $\epsilon \le \eta$, any tour $\ket{T}$ having $\phi(T) = 2\pi-\epsilon/2$ is also regarded as an approximate solution, defined as Eq.~(\ref{eq:conda1}). Fortunately, we can exclude such highest-cost tours by using ``quantum phase estimation'' within a small number of additional queries. For quantum phase estimation, the ancillary register is coupled with the register for eigenstates $\ket{T}$ of $\hat{C}$ with their eigenvalues $e^{i\phi(T)}$. If size of the ancillary register is $m$-qubit, we can estimate a $\phi(T)$ within $2^m$ queries \cite{Nielsen99,Dobsicek07}. Here, estimation error $\delta$ is given by $\delta \le 1/2^{m}$. If $\delta \le {\Delta\phi}$ taking into account the action of $\hat{C}$, the best estimation is done within ${\cal O}(1/{\Delta\phi}) = {\cal O}(n^{1/2})$ queries. Therefore, the complexity of the algorithm is essentially unchanged and quadratic speedup is retained.

\section{Summary}

As the classical heuristic algorithm is much faster than the exact quantum algorithm, we proposed a quantum heuristic algorithm to solve TSP by generalizing the structure of Grover search. The generalized operation consists of the conventional reflection with respect to the equal superposition state and the cost oracle which answers the cost phase when queried, distinct from the previous generalizations. Sufficient conditions were derived to greatly enhance the probability of finding the cheapest tours and they were characterized by the statistical properties of tour costs. In particular for the Gaussian distribution of the tours along the cost, we showed that the conditions are satisfied and the proposed quantum heuristic algorithm exhibits the quadratic speedup of its classical heuristic algorithm. It is thus an open question whether the quadratic or faster speedup can still be achieved when employing the structural or geometrical information of the city-pair costs.

\end{document}